\documentclass[final,3p,times]{elsarticle}

\usepackage{amssymb}
\usepackage{amsmath}
\usepackage{graphicx}
\usepackage{bm}
\usepackage{float}
\usepackage[english]{babel}
\usepackage{arydshln}
\usepackage{url}

\journal{Journal of Computational Physics}

\begin{document}

\begin{frontmatter}



\title{The compressible Neural Particle Method for Simulating Compressible Viscous Fluid Flows}


\author[label1]{Masato Shibukawa} 
\author[label2]{Naoya Ozaki}
\author[label3]{Maximilien Berthet}

\affiliation[label1]{organization={Space and Astronautical Science, The Graduate University for Advanced Studies, SOKENDAI},
            addressline={Syohankokusai-mura, Hayama-chou}, 
            city={Miura-gun},
            postcode={240-0015}, 
            state={Kanagawa},
            country={Japan}}
\affiliation[label2]{organization={JAXA/ISAS},
            addressline={3-1-1, Yuranodai, Chuou-ku}, 
            city={Sagamihara-shi},
            postcode={252-5210}, 
            state={Kanagawa},
            country={Japan}}
\affiliation[label3]{organization={Department of Aeronautics and Astronautics, Graduate School of Engineering, University of Tokyo},
            addressline={5-1-5, Kashiwanoha}, 
            city={Kashiwa-shi},
            postcode={270-0882}, 
            state={Chiba},
            country={Japan}}
\begin{abstract}
Particle methods play an important role in computational fluid dynamics, but they are among the most difficult to implement and solve. The most common method is smoothed particle hydrodynamics, which is suitable for problem settings that involve large deformations, such as tsunamis and dam breaking. However, the calculation can become unstable depending on the distribution of particles. In contrast, the neural particle method has high computational stability for various particle distributions is a machine learning method that approximates velocity and pressure in a spatial domain using neural networks. The neural particle method has been extended to viscous flows, but until now it has been limited to incompressible flows. In this paper, we propose the compressible neural particle method, which is a new feed-forward neural network-based method that extends the original neural particle method to model compressible viscous fluid flows. The proposed method uses neural networks to calculate the velocity and pressure of fluid particles at the next time step, and the Tait equation to calculate the density to handle the compressibility. The loss function is composed of the governing equations of compressible flow and the boundary conditions, which are free surface and solid boundary conditions. We demonstrate that the proposed method can accurately solve the compressible viscous fluid flow, a problem that was difficult to solve with the smoothed particle hydrodynamics method, by applying it to a dam breaking problem.
\end{abstract}

\begin{keyword}


SPH \sep Compressible flow \sep Physics Informed Neural networks \sep Machine Learning \sep Neural Particle Method
\end{keyword}

\end{frontmatter}


\section{Introduction}
\label{sec1}
Particle methods are numerical techniques widely used in computational fluid dynamics (CFD) to analyze fluid motion and heat transfer \cite{Lind2020} in various fields such as aerospace, energy, and health care \cite{Shadloo2016}. Among these methods, the Smoothed Particle Hydrodynamics (SPH) technique is the most widely used, especially for simulations involving large deformations \cite{Ye2019}. SPH is a mesh-free, Lagrangian method in which the fluid is modeled as a collection of particles that carry mass, momentum, and energy. The field variables are computed by smoothing, or kernel-weighted averaging, over neighboring particles. This approach enables SPH to naturally handle complex, highly deformable, and free-surface flows without requiring a fixed computational grid. However, the stability of SPH calculations in CFD can be affected by the distribution of particles. 

In particular, particle clustering and local particle deficiency often result in non-physical pressure fluctuations and numerical divergence \cite{lind2012}. To address these issues, various methods have been proposed in recent years. One such approach is the particle shifting technique, which reduces particle distribution bias by periodically repositioning particles by a small amount to maintain uniformity \cite{xu2009, sun2012}. Xu \cite{xu2009} proposed a method that slightly shifts particles away from streamlines to achieve a more uniform distribution, addressing the problem of particle concentration in certain regions that causes computational instability. Sun \cite{sun2012} introduced a physics-based particle transport model utilizing Fick's diffusion law, improving numerical stability in a wider range of cases. However, particle repositioning introduces excessive numerical dissipation, which violates physical laws, and it remains challenging to appropriately set parameters such as the magnitude and frequency of repositioning. Another approach is to adopt an adaptive kernel, which dynamically adjusts the kernel size based on local particle density \cite{vacondio2013}. Nevertheless, this method faces limitations: increasing the kernel size raises computational costs, and abrupt changes in kernel size may fail to adapt to varying particle densities.

Additionally, modifications to the fundamental SPH equations have been actively studied \cite{molteni2009, colagrossi2003, marrone2011}. Molteni \cite{molteni2009} extended the conventional artificial viscosity by adding a new density diffusion term to the mass conservation equation. Colagrossi \cite{colagrossi2003} proposed a correction term in SPH to handle two-dimensional interface flows with low density ratios. Marrone \cite{marrone2011} developed the $\delta$-SPH method, incorporating numerical diffusion terms into the density field. While these methods have mitigated numerical instability, artificial dissipation and sensitivity to parameter settings remain significant challenges. Moreover, Hopkins \cite{hopkins2015} proposed a hybrid method combining SPH with grid-based approaches, which demonstrated high effectiveness in stabilizing particle distributions. However, the complexity of hybrid algorithms and the associated increase in computational cost have emerged as new issues. In summary, improving SPH remains an open challenge, characterized by non-physical processes, high computational costs, sensitivity to parameter settings, and difficulties in algorithm development.

In recent years, various approaches have been proposed to integrate deep learning into computational fluid dynamics (CFD) \cite{Kutz2017, Kochkov2021}.
Among these, physics-informed neural networks (PINNs) have shown considerable promise \cite{Raissi2019, Cai2022}.
Unlike conventional data-driven models, PINNs incorporate physical laws—typically formulated as partial differential equations (PDEs)—into the training process.
This allows accurate modeling of physical systems even with limited data, by simultaneously minimizing the residuals of both observational data and governing equations \cite{Raissi2019}.
PINNs have been applied not only to forward modeling \cite{Cai2022}, but also to inverse problems, such as inferring hidden variables from sparse measurements \cite{Mao2020}.
Recent advancements have addressed some critical limitations of standard PINNs. Ko \cite{Ko2025} proposed a variable-scaling approach (VS-PINN) that significantly enhances training stability and accuracy when solving stiff PDEs.
Yadav \cite{Yadav2024} introduced RF-PINNs, a framework for reconstructing full reactive flow fields using only sparse velocity or temperature data, demonstrating effective generalization across laminar and turbulent flame regimes.
In parallel, researchers have investigated the application of deep learning to particle-based methods in fluid dynamics \cite{Jin2021, Raissi2020, Haghighat2021}.
One notable development is the Neural Particle Method (NPM) \cite{Wessels2020}, a mesh-free framework that integrates PINNs into particle simulations, providing robust performance across complex flow scenarios.

The NPM \cite{Wessels2020} applies PINNs to particle-based modeling of inviscid incompressible fluid dynamics. It discretizes time using the implicit Runge-Kutta (IRK) method and computes the velocity and pressure of particles at each time step using a neural network. This network incorporates PINNs to approximate the relationships among position, velocity, and pressure while satisfying the laws of momentum and mass conservation. Moreover, NPM introduces a distance function and a penalty term to strictly enforce Dirichlet boundary conditions and prevent particles from penetrating solid walls. NPM has shown excellent agreement with experimental results in dam-break problems. To improve the applicability of NPM to viscous flows, Jin proposed the General Neural Particle Method (gNPM) \cite{Bai2022}. By incorporating the viscosity term into the momentum conservation law, gNPM enables modeling of viscous fluid dynamics. Furthermore, gNPM enhances computational efficiency by eliminating redundant processes in NPM. In NPM, time discretization using the IRK method requires multiple velocity and pressure calculations for each particle at every time step, based on the IRK order. Since the governing equations include the time derivative of velocity but not of pressure, pressure is predicted by computing the weighted average of multiple pressure outputs. gNPM simplifies this process by outputting a single pressure value, thereby reducing the neural network size and lowering computational costs. gNPM achieves high computational stability even with an arbitrary particle distribution, thanks to the strong spatial approximation capability of the neural network. This stability is maintained without additional non-physical corrections, such as compensation terms or particle displacement algorithms, which conventional SPH methods rely on. Extensive simulations under various conditions, including sloshing, Couette flow, and dam-break scenarios, have demonstrated the high practicality of gNPM. Moreover, gNPM successfully performs simulations even in cases where traditional SPH fails, for example when randomly initialized particle positions lead to numerical instability.These results confirm its high computational stability. However, gNPM cannot handle compressible fluids and does not compute density variations.

In this study, we propose an extended neural particle method that incorporates density variations to overcome the limitations of gNPM. By integrating the equation of state and continuity equation into the neural network framework, the proposed method can model compressible fluid dynamics while maintaining the advantages of gNPM, such as stability and accuracy in particle distribution. Additionally, we improve the time integration scheme and neural network architecture to reduce computational cost and enhance scalability. Numerical experiments under various conditions demonstrate that the proposed method can successfully handle compressible fluid simulations that are difficult to model with conventional SPH or gNPM approaches. These results indicate the potential of neural network-based particle methods as a promising alternative for simulating complex fluid phenomena.

\section{Govering equations of fluid dynamics}
\label{sec3}
The governing equations of fluid dynamics for compressible viscous flow are based on the conservation laws of mass and momentum. The equations are written in the Lagrangian formulation as follows:
\begin{equation}
\frac{\partial \rho(\bm{x}, t)}{\partial t} = -\rho(\bm{x}, t) \nabla \cdot \bm{v}(\bm{x}, t), \quad
\frac{\partial \bm{v}(\bm{x}, t)}{\partial t} = -\frac{1}{\rho(\bm{x}, t)} \nabla p(\bm{x}, t) + \frac{1}{\rho(\bm{x}, t)} \nabla \cdot (\mu \nabla \bm{v}(\bm{x}, t)) + \bm{g}
\label{eq:gov_eq}
\end{equation}
where \( \bm{v} \) and \( p \) represent the fluid velocity and pressure respectively. \( \rho \) is the fluid density, and \( \mu \) is the dynamic viscosity coefficient. \( \bm{g} \) refers to gravitational acceleration. For weakly compressible fluids where density variations are within a few percent, the relationship between pressure and density can be described using the Tait equation \cite{Becker2007, Monaghan1994}. In the cNPM, the density is to be calculated from the pressure, so the Tait equation is written as follows:
\begin{equation}
\rho = \rho_0 \left( \frac{p}{B} + 1 \right)^{\frac{1}{\gamma}}, \quad B = \frac{\rho_0 c_s^2}{\gamma}, \quad c_s = \frac{v_f}{\sqrt{\eta}}
\label{eq:tait}
\end{equation}
where $p$ and $\rho_0$ represent the fluid pressure and the initial fluid density, respectively. \( \gamma \) remains constant, we set $\gamma = 7$. $B$ governs the rate of change of density. $c_s$ refers to the speed of sound. $\eta$ remains constant, we set $\eta=0.01$. $v_f$ is the maximum expected velocity of the fluid under the simulation conditions\cite{Becker2007}. In this study, we use the Tait equation, but it is possible to extend it to other equations of state easily. 

\section{Compressible Neural Particle Method}
\label{sec3}
\begin{figure}[t]
  \centering
  \includegraphics[width=\textwidth]{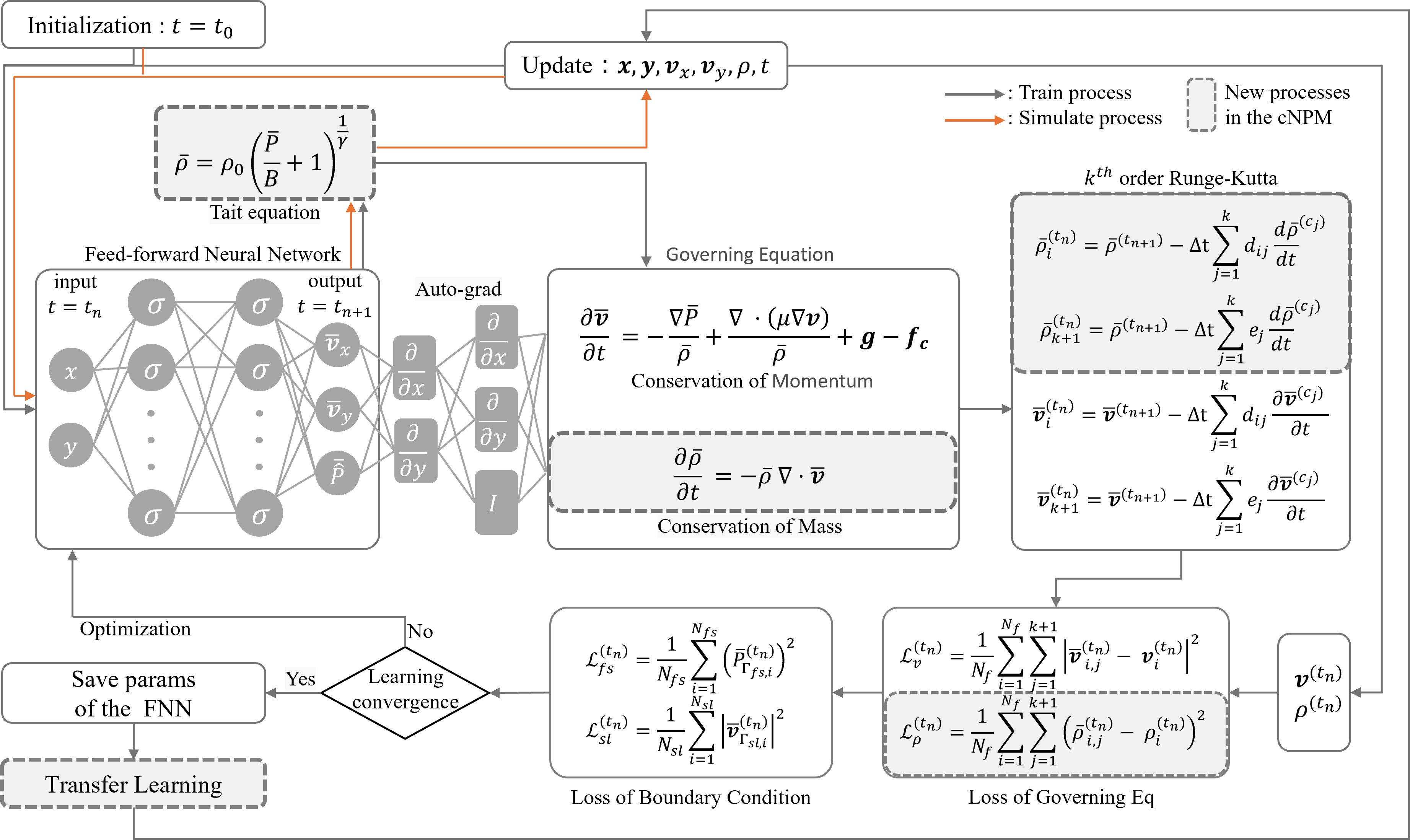} 
  \caption{The flowchart of the compressible Neural Particle Method (the cNPM).} 
  \label{fig:flowchart} 
\end{figure}

This section explains the compressible Neural Particle Method (cNPM), which combines the state equation with the NPM framework. In the cNPM, we first discretize the time domain. At each time step, a Feed-forward Neural Network (FNN) \cite{Schmidhuber2015} learns and computes the velocity and pressure for the next time step \cite{Wessels2020}. We calculate the particle positions and density analytically from the equation of motion and the state equation, respectively. The fluid motion dynamics consist of many particles. Figure \ref{fig:flowchart} shows the flowchart of the cNPM.

Let us explain the computation procedure of the cNPM at time $t_n$. We input the positions to the FNN and obtain the velocity and pressure for the next time step as outputs. Automatic differentiation allows us to compute the partial derivatives of the velocity and pressure. The Tait equation provides the density. To predict the density and velocity at the previous time step, the cNPM uses the governing equations and the (Implicit or Explicit) RK method. The loss function incorporates the boundary conditions, which consist of solid and free surface conditions, and the governing equations (Eq.\ref{eq:gov_eq}). We train the FNN to minimize the loss function through error backpropagation. We determine the initial parameters of the FNN by transfer learning, using the parameters from the previous time step. Once this error falls below a specified threshold, we save the parameters of the FNN and use them for transfer learning. The cNPM simulates and updates the particle positions $x$, $y$ from the velocity $v_x$, $v_y$. This procedure repeats until the final time step.

\subsection{Feed-forward Neural Network}
\label{sec3.1}
The FNN \cite{Schmidhuber2015} refers to a method of approximating a nonlinear space by arranging a large number of neurons in multiple layers. Figure~1 (upper left) illustrates the conceptual diagram of the FNN. The input is processed by accumulated weights, added biases, and then substituted into a nonlinear function named activation function. This calculation is repeated layer by layer until the output is produced. This procedure is represented as follows:
\begin{equation}
\bm{h}^{(l)} = \sigma(\bm{W}^{(l)} \bm{h}^{(l-1)} + \bm{b}^{(l)}) \quad l = 1, \ldots , L
\label{eq:fnn_layer}
\end{equation}
Where \( \bm{h}^{(l)} \) denotes the output of the \( l \)-th layer, \( \sigma \) is the activation function, \( \bm{W}^{(l)} \) is the weight matrix, and \( \bm{b}^{(l)} \) is the bias vector. In the 1st layer, we use the particle position as \( \bm{h}^{(1)} \), which serves as the input. In the \( L \)-th layer, we get \( \bm{h}^{(L)} \) as velocity and pressure, which are the outputs. During the FNN training, the FNN optimizes the weights \( \bm{W}^{(l)} \) and biases \( \bm{b}^{(l)} \) to minimize the error between the network prediction and the true value. The outputs of the FNN are the velocity and pressure because the governing equations (Eq. \ref{eq:gov_eq}) require the gradients of velocity and pressure, which we can easily compute using automatic differentiation\cite{Baydin2018}. We obtain the density via the Tait equation (Eq. \ref{eq:tait}) from the predicted pressure. Based on these considerations, we describe the FNN model as follows:
\begin{equation}
v_x^{(t_{n+1})}, v_y^{(t_{n+1})}, p^{(t_{n+1})} = FNN(x^{(t_n)}, y^{(t_n)})
\label{eq:fnn}
\end{equation}
Where \( x, y \) represent the particle position, \( v_x, v_y \) denote the velocity in the \( x \)- and \( y \)-directions, respectively, \( p \) is the pressure, and \( FNN(x, y) \) denotes the FNN as a function. We discretize the time domain as \( t_n = t_0 + n \Delta t (n = 0, 1, 2, \ldots) \). \(\Delta t\) is the time increment. We compute the state of the particle ensemble at each time \( t \) by training the FNN step by step. In the cNPM, we do not automatically adjust the time increment, but keep it fixed for all steps.
\subsection{Loss Function}
\label{sec3.3}
The errors to be minimized in cNPM stem from the governing equations of compressible viscous fluid flow and the boundary conditions. We define the error in the governing equations as the difference between the left and right sides of Eq.\ref{eq:gov_eq}. This error becomes close to zero as training progresses, indicating that the fluid field (spatial distribution of velocity, pressure, and density) satisfies the equations. To evaluate the right-hand side, we need the gradients of velocity and pressure. However, it is difficult for the FNN to directly handle the time evolution on the left-hand (i.e., the time derivatives of density and velocity). The FNN does not explicitly take time as an input and cannot differentiate with respect to time by using automatic differentiation. Therefore, we propagate the predicted state at time \( t_{n+1} \) backwards to estimate the state at time \( t_n \), and use the difference from the state at time \( t_n \) (which has already converged through training) to evaluate the error indirectly. We perform this propagation using the Runge-Kutta (RK) method, which combines implicit and explicit approaches\cite{Bai2022}. A \( k \)-th order Runge-Kutta method accurately propagates the solution in time by taking a weighted average of \( k \) predicted temporal changes\cite{Iserles2009}. To use the \( k \)-th RK method, the FNN outputs \( k+1 \) velocities for the same position. The backward propagation is expressed by the following equation:
\begin{equation}
\begin{aligned}
\bar{\rho}_i^{(t_n)} &= \bar{\rho}^{(t_{n+1})} - \Delta t \sum_{j=1}^k d_{ij} \left( \frac{d\bar{\rho}}{dt} \right)^{(c_j)}, \quad i=1,\ldots,k \\
\bar{\rho}_{k+1}^{(t_n)} &= \bar{\rho}^{(t_{n+1})} - \Delta t \sum_{j=1}^k e_{j} \left( \frac{d\bar{\rho}}{dt} \right)^{(c_j)} \\
\bar{\bm{v}}_i^{(t_n)} &= \bar{\bm{v}}^{(t_{n+1})} - \Delta t \sum_{j=1}^k d_{ij} \left( \frac{\partial \bar{\bm{v}}}{\partial t} \right)^{(c_j)}, \quad i=1,\ldots,k \\
\bar{\bm{v}}_{k+1}^{(t_n)} &= \bar{\bm{v}}^{(t_{n+1})} - \Delta t \sum_{j=1}^k e_{j} \left( \frac{\partial \bar{\bm{v}}}{\partial t} \right)^{(c_j)}
\end{aligned}
\end{equation}
Where \( \bar{\rho}_i^{(t_n)} \) and \( \bar{\bm{v}}_i^{(t_n)} \) denote the density and velocity at time step \(t_n\), respectively. \( d_{ij} \) and \(e_{i} \) are coefficients of the RK method, their indices \(i\) and \(j\) are the RK stage. At the time \( t \), a converged solution, which is considered correct, has already been obtained, so the mean square error of velocity and density is calculated on the basis of this solution. The loss of the governing equations is then defined as follows:
\begin{equation}
  \begin{aligned}
  \mathcal{L}_v^{(t_n)} &= \frac{1}{N_f}\sum_{i=1}^{N_f} \sum_{j=1}^{k+1} |\bar{\bm{v}}_{i,j}^{(t_n)} - \bm{v}_{i}^{(t_n)}|^2 \\
  \mathcal{L}_{\rho}^{(t_n)} &= \frac{1}{N_f} \sum_{i=1}^{N_f} \sum_{j=1}^{k+1} (\bar{\rho}_{i,j}^{(t_n)} - \rho_{i}^{(t_n)})^2 
  \end{aligned}
  \end{equation}
where \( \mathcal{L}_v^{(t_n)} \) and \( \mathcal{L}_{\rho}^{(t_n)} \) represents the loss function associated with the velocity and density at time step \(t_n\), respectively. \( N_f \) is the total number of fluid particles. 

The boundary condition errors include contributions from the free surface boundary condition and the solid boundary condition based on the Dirichlet condition. The pressure at the free surface is zero because the particles are in contact with the atmosphere. The velocity of the solid particles forming walls or floors is also zero at all time steps. It should be noted that, since cNPM does not include an algorithm to dynamically identify particles on the free surface, these particles remain fixed on the surface throughout all time steps. As a result, the predicted values of these particles directly represent the boundary errors, and we use them to compute the loss function as follows:
\begin{equation}
  \begin{aligned}
    \mathcal{L}_{fs}^{(t_n)} &= \frac{1}{N_{fs}} \sum_{i=1}^{N_{fs}} \left( \bar{p}_{\Gamma_{fs,i}}^{(t_n)} \right)^2\\
    \mathcal{L}_{sl}^{(t_n)} &= \frac{1}{N_{sl}}\sum_{i=1}^{N_{sl}} |\bar{\bm{v}}_{\Gamma_{sl,i} }^{(t_n)}|^2 \\
  \end{aligned}
\end{equation}
where \( \mathcal{L}_{fs}^{(t_n)} \) and \( \mathcal{L}_{sl}^{(t_n)} \) represent the loss functions associated with the free surface and solid boundary conditions at time step \( t_n \), and \( N_{fs} \) and \( N_{sl} \) are the total number of free surface and solid particles, respectively. \( \bar{p}_{\Gamma_{fs,i}}^{(t_n)} \) is the predicted pressure at the free surface, and \( \bar{\bm{v}}_{\Gamma_{sl,i}}^{(t_n)} \) is the predicted velocity at the solid boundary. In addition, we introduce a penalty term that uses a spring force to prevent particles from entering the solid region. We add a repulsive force to the equations of motion that depends on the distance between the particles and the solid \cite{Bai2022}. The penalty term is written as follows:
\begin{equation}
  \bm{f}_c^i=\varepsilon_c g^i a\left(g^i\right) \bm{n}, \quad g^i=\left(\bm{x}^i-\overline{\bm{x}}\right) \cdot \bm{n}, \quad a\left(g^i\right)=0.5\left(1+\operatorname{sign}\left(g^i\right)\right)= \begin{cases}0 & \text { if } g^i<0 \\ 0.5 & \text { if } g^i=0 \\ 1 & \text { if } g^i>0\end{cases}
\end{equation}
where \( \bm{f}_c^i \) is the penalty force acting on the \( i \)-th particle, \( \varepsilon_c \) is the penalty coefficient, \( \bm{n} \) is the normal vector of the solid boundary, \( \bm{x}^i \) is the position of the \( i \)-th particle, and \( \overline{\bm{x}} \) is the position of the solid boundary. We obtain the total loss function \( \mathcal{L} \) by summing the governing equation errors and the boundary condition errors, and express it by the following equation.
\begin{equation}
\mathcal{L} = \mathcal{L}_v + \mathcal{L}_{\rho} + \mathcal{L}_{fs} + \mathcal{L}_{sl}
\end{equation}
When we include different physical quantities in the loss function, we need to balance the orders of magnitude to avoid instability in the learning process. Because the orders of magnitude of pressure and density are about 1000 times larger than those of velocities, we scale pressure as \( p = \rho_0 \hat{p} \) and density as \( \rho = \rho_0 \hat{\rho} \). \( \hat{\rho} \) and \( \hat{p} \) are the scaled density and pressure, respectively, and \( \rho_0 \) denotes the initial density. 
The factor of 1000 used here is specific to the dam-break problem analyzed later in this study, as it reflects the typical ratio of pressure and density to velocity in that particular flow. However, the approach of scaling pressure and density relative to a characteristic velocity or reference density is a general technique that can be applied to other problems as well. Under these conditions, we transform Eqs.~\ref{eq:gov_eq} and \ref{eq:tait} as follows:
\begin{equation}
\frac{\partial \hat{\rho}}{\partial t} = -\hat{\rho} \nabla \cdot \bm{v}, \quad \frac{\partial \bm{v}}{\partial t} = -\frac{\nabla \hat{p}}{\hat{\rho}} + \frac{\nabla \cdot (\mu \nabla \bm{v})}{\rho_0 \hat{\rho}} + \bm{g}
\end{equation}
\begin{equation}
\hat{\rho} = \left( \frac{\gamma \hat{p}}{c_s^2} + 1 \right)^{\frac{1}{\gamma}}
\end{equation}
The governing equations, discretized in the temporal direction, are formulated based on the Updated Lagrangian formulation. In these equations, the terms $\nabla \cdot \bm{v}$ and $\nabla \hat{p}$ are defined following the formulation proposed by Wessels\cite{Wessels2020}. Furthermore, the viscous term, $\nabla \cdot (\mu \nabla \bm{v})$, is detailed in the Appendix section.
\subsection{Transfer Learning}
\label{sec3.4}
We apply transfer learning to the FNN trained on one task and use it for another, related task\cite{Zhuang2021}. This method can significantly reduce the training time if the new task is similar to the original or if the original task is large. In cNPM, we train at every time step, but large variations in velocity, pressure, and density fields are unlikely to occur between small time increments. Therefore, we save the parameters of the FNN after training at time \( t_n \) and use them as initial parameters for the next time step \( t_{n+1} \).

\section{Numerical Examples}
The compressible Neural Particle Method (cNPM) was applied to simulate a dam breaking problem assuming compressible viscous flow of water. The parameters of the FNN were determined through a systematic grid search over the network architecture (e.g., number of layers and neurons) and the learning rate, based on the considerations above. The cNPM was developed using Pytorch and simulated in a 64-bit Linux environment on a system consisting of an Intel(R) Core(TM) i9-10980XE CPU and an NVIDIA Quadro GP100 GPU.

\begin{table}[H]
  \centering
  \caption{Experimental Configuration}
  \begin{tabular}{lll}
  \hline
  \textbf{Name} & \textbf{Value} & \textbf{Unit/Remarks} \\
  \hline
  \textbf{Physics parameter} & & \\
  Fluid & Water &  \\
  Dynamic Viscosity Coefficient & $1.0 \times 10^{-3}$ & Ns/m$^2$ \\
  Initial Density & $997$ & kg/m$^3$ \\
  Gravitational Acceleration & 9.81 & m/s$^2$ \\
  Dam Height & 0.2 & m \\
  Dam Length & 0.1 & m \\
  \hdashline
  \textbf{Tait equation parameter} & & \\
  Maximum Fluid Velocity ($v_f$) & 1.98 & m/s \\
  Constant $\eta$ & 0.01 &  \\
  Speed of Sound ($c_s$) & 19.8 & m/s \\
  Tait Equation Coefficient ($B$) & $5.58 \times 10^4$ & Pa, $\gamma = 7$ \\
  \hdashline
  \textbf{FNN parameter} & & \\
  Structure & 4 layers, 50 neurons per hidden layer &  \\
  Optimizer & Adam &  \\
  Learning Rate & $1.0 \times 10^{-2}$ & \\
  Activation Function & Mish\cite{Misra2019} &  \\
  \hdashline
  \textbf{Other parameter} & & \\
  Total Simulation Particles & 64,800 &  \\
  Training Particles & 800 &  \\
  Particle Arrangement & 20 (horizontal) $\times$ 40 (vertical) &  \\
  Time Step Size & 0.005 & s \\
  Total Simulation Time & 0.2 & s \\
  \hline
  \end{tabular}
  \label{tab:experiment}
\end{table}

\begin{figure}[H]
  \centering
  \includegraphics[width=0.8\textwidth]{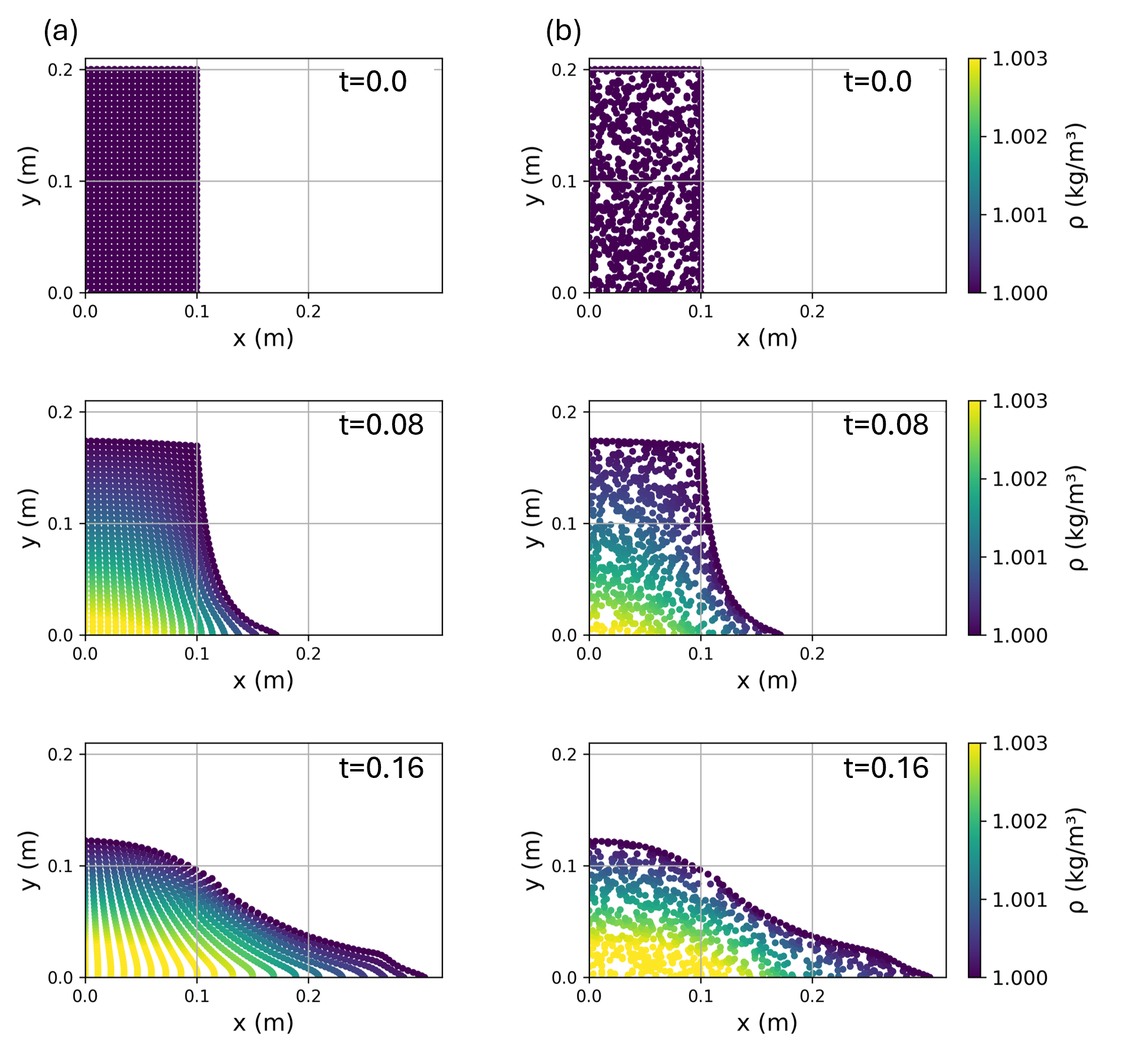}
  \caption{Temporal evolution of the dam shape and pressure distribution. Rows correspond to time steps at \(t = 0\), \(t = 0.08\,\mathrm{s}\), and \(t = 0.5\,\mathrm{s}\). The left-hand plots display results obtained with a uniform initial particle distribution, while the right-hand plots show the results using a randomized initial particle distribution.
}
  \label{fig:press_dis} 
\end{figure}

The cNPM method was systematically evaluated under two distinct initial particle configurations: a uniform grid and a randomized distribution. 
Figure~\ref{fig:press_dis} illustrates the resulting density contours at \(t = 0\), \(0.08\,\mathrm{s}\), and \(0.16\,\mathrm{s}\) for both cases. 
Remarkably, cNPM consistently yields a smooth and physically coherent density field across all scenarios, even in the presence of randomized particle arrangements that introduce local voids.
No spurious spikes or filament-like artifacts are observed, underscoring the method’s stability and accuracy.
Notably, the density contours in the randomized case closely align with those of the uniform grid, highlighting cNPM’s inherent robustness to spatial irregularity.
This distribution-invariant behavior starkly contrasts with conventional SPH-based approaches, which frequently exhibit significant noise and instability in the presence of voids. 
The persistent similarity between the two configurations—uniform and randomized—suggests that cNPM effectively exploits collective particle dynamics to maintain field continuity.
The free surface remains smooth and unperturbed throughout the simulation, further affirming the method’s resilience to randomness-induced distortions. 
From a physical standpoint, density values near the bottom boundary and the lower-left solid wall converge to approximately \(\rho \approx 1.003\), whereas the density near the free surface remains around \(\rho \approx 1.000\). 
The maximum relative density deviation is less than \(1\%\), aligning with the compressibility constraint imposed by the Tait equation. 
This indicates that the selected Tait parameter appropriately regulates density evolution, yielding physically meaningful and dynamically stable behavior.
Boundary condition implementation was verified by confirming that (1) particles do not penetrate the solid boundaries, and (2) the free-surface density consistently remains near \(\rho = 1.000\), corresponding to an effectively zero pressure under the Tait formulation. 

\begin{figure}[H]
  \centering
  \includegraphics[width=\textwidth]{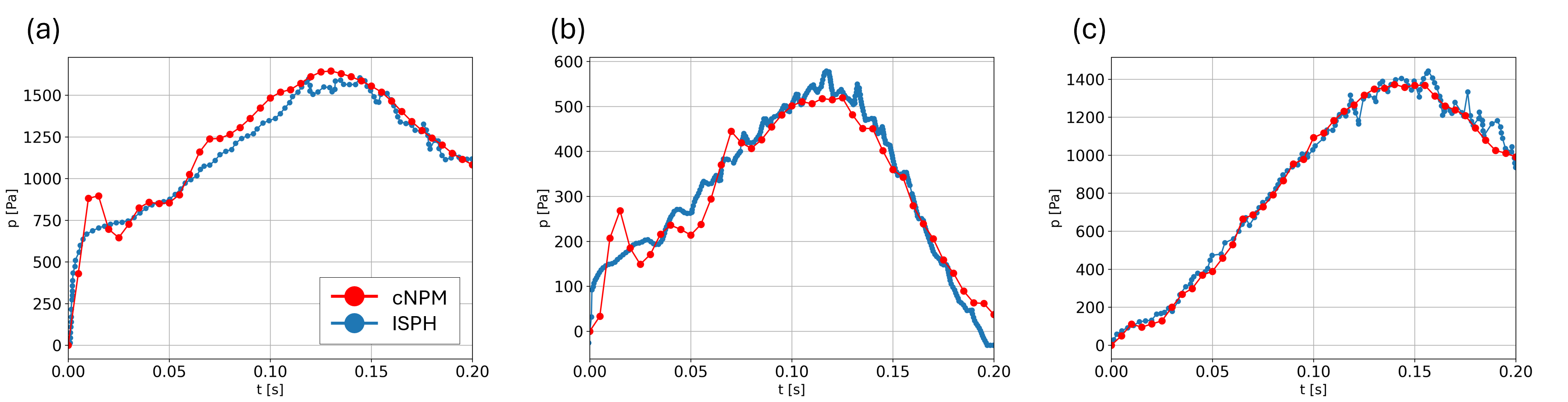}
  \caption{Pressure–time histories computed by cNPM and iSPH\cite{Bai2022} at three wall-mounted monitoring points: \((0,0)\) (left), \((0.1,0)\) (center), and \((0,0.1)\) (right).
}
  \label{fig:pressure_compare} 
\end{figure}

Figure~\ref{fig:pressure_compare} illustrates the time evolution of pressure at three distinct monitoring locations on the wall—\((0,0)\), \((0.1, 0)\), and \((0, 0.1)\)—as computed by both cNPM and implicit SPH (iSPH)~\cite{Bai2022}. The initial distribution of particles is a regular grid configuration. At all three points, the rise time, peak magnitude, and subsequent decay of the pressure signals exhibit close agreement between the two methods.
A minor discrepancy is observed at \((0, 0)\) and \((0.1, 0)\) around \(t \approx 0.025\,\mathrm{s}\), where cNPM predicts an elevated pressure peak compared to iSPH. 
Despite this deviation, the overall pressure trends remain consistent across both approaches.
These results demonstrate that cNPM provides reliable predictions of transient pressure loads on solid boundaries. 

\begin{figure}[H]
  \centering
  \includegraphics[width=0.8\textwidth]{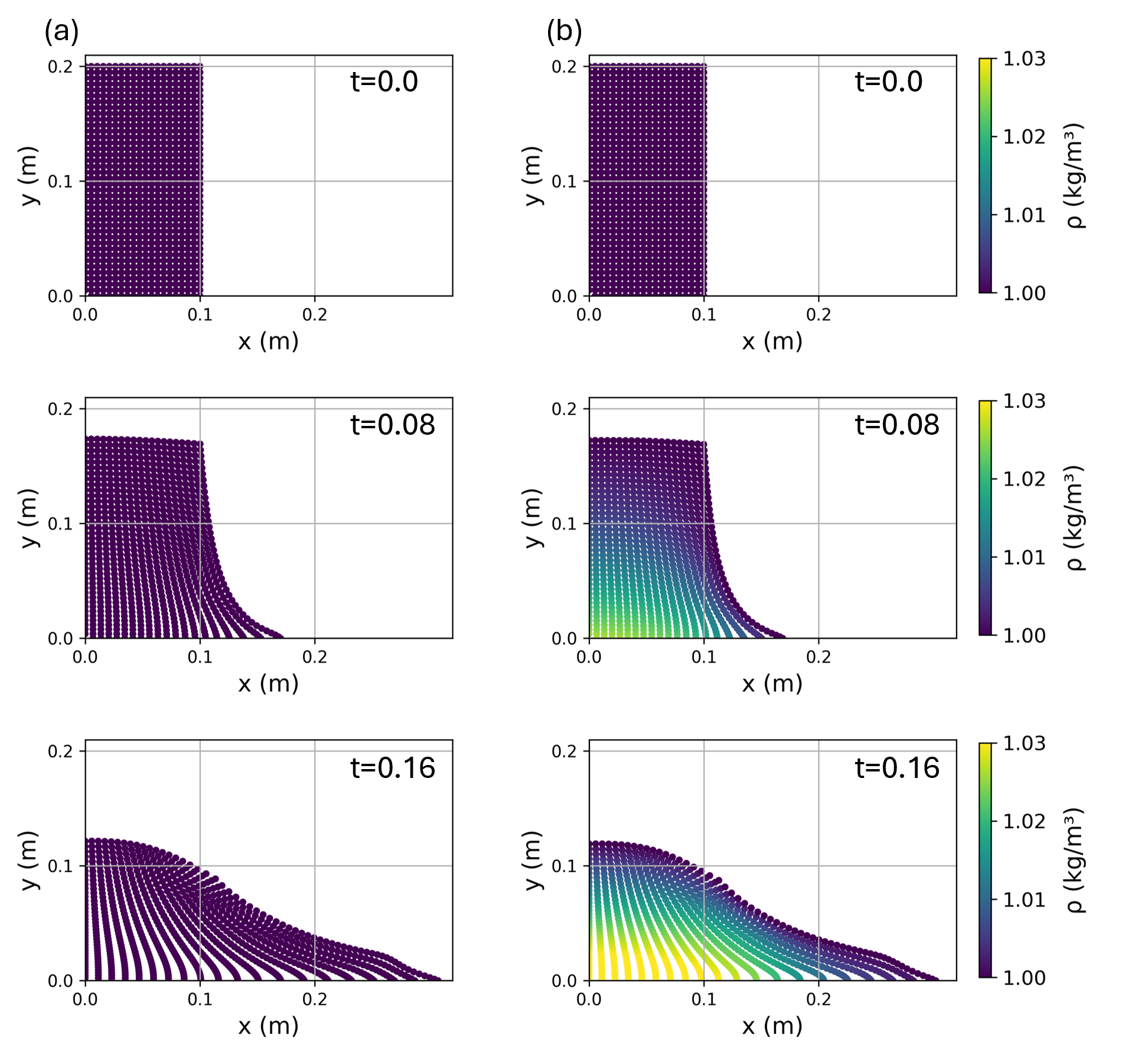}
  \caption{This figure illustrates the temporal evolution of the dam shape and density distribution. Columns correspond to time steps at \(t = 0\), \(t = 0.08\,\mathrm{s}\), and \(t = 0.16\,\mathrm{s}\). The left line displays results obtained with a nearly incompressible setting with \(\eta = 0.001\), while the right line shows the results using a highly compressible configuration with \(\eta = 1\).
  }
  \label{fig:press_dis_params}
\end{figure}

A parametric study was performed to evaluate the influence of the density-change tolerance coefficient \(\eta\) in the Tait equation on simulation outcomes. Two cases were compared: a nearly incompressible setting with \(\eta = 0.001\), and a highly compressible configuration with \(\eta = 1\).
Figure~\ref{fig:press_dis_params} (left) presents density contours at \(t = 0.00\), \(0.08\,\mathrm{s}\), and \(0.16\,\mathrm{s}\) under the strict tolerance of \(\eta = 0.001\). In this setting, density variations remain negligible throughout the simulation, indicating fluid behavior that closely approximates incompressibility.
Figure~\ref{fig:press_dis_params} (right) shows results for \(\eta = 1\), using a color scale adjusted to reflect a tenfold increase in allowable density variation relative to the \(\eta = 0.01\) case in Figure~\ref{fig:press_dis}. Under this more compressible regime, higher localized density emerges in high-pressure regions, particularly near the bottom boundary. Nevertheless, the overall dam-front propagation and free-surface evolution remain visually consistent with those of the nearly incompressible case.
While slight deviations in free-surface profile and leading-edge position are observable—primarily at the scale of individual particle spacing—the global flow characteristics are effectively unchanged. These findings suggest that variations in the Tait coefficient have only a limited impact on the dam-break kinematics in the present scenario.

\begin{figure}[H]
  \centering
  \includegraphics[width=0.8\textwidth]{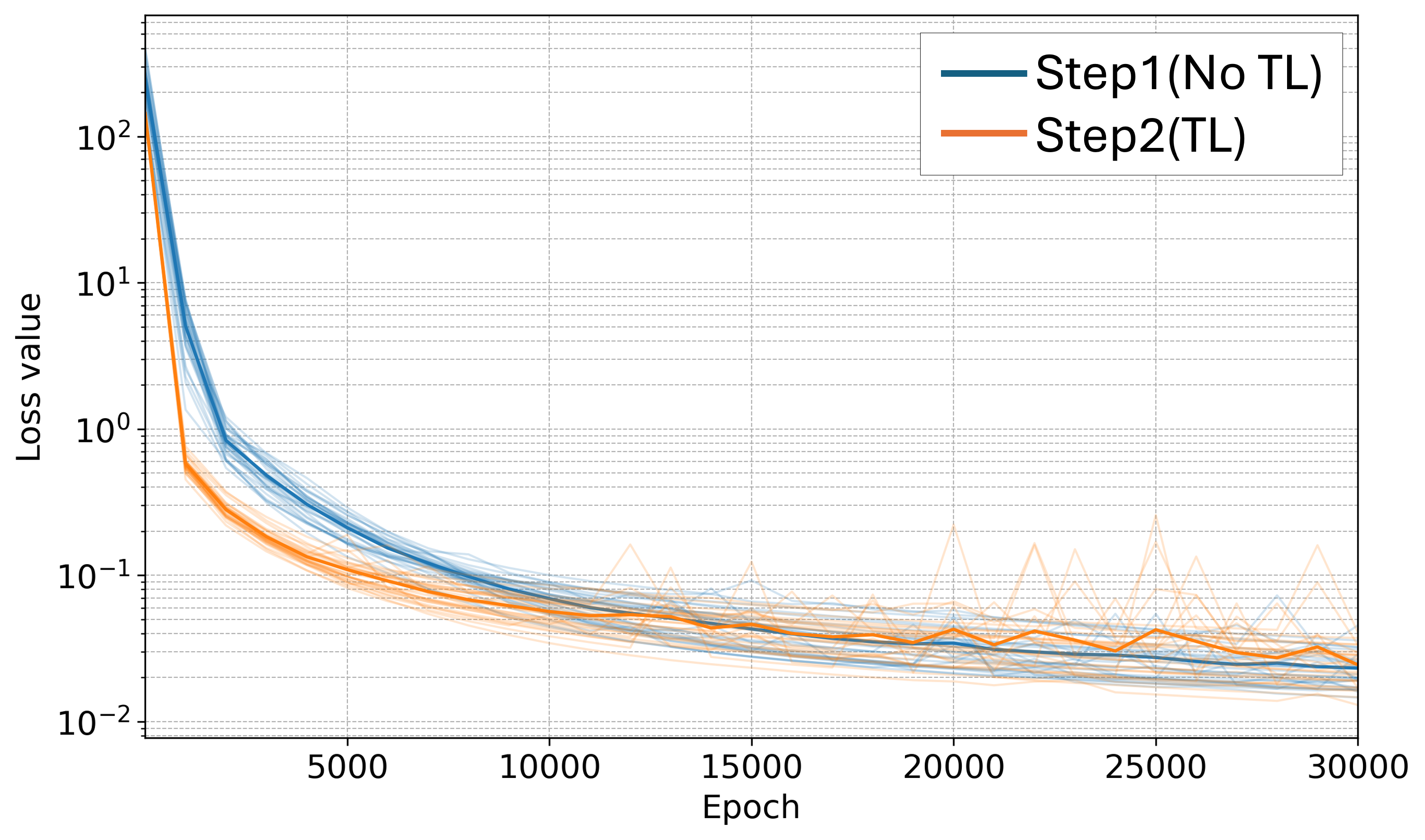}
  \caption{Loss curves over 30{,}000 epochs for training with (orange) and without (blue) transfer learning. Thin lines show individual runs (25 each), and thick lines indicate the average loss.
  }
  \label{fig:loss_curve}
\end{figure}

In order to evaluate the training efficiency through transfer learning, we compare two-stage training scenarios using the cNPM model: the 1st time step, which begins with randomly initialized weights, and the 2nd time step, which uses the trained weights from the 1st time step as initialization. Each stage was repeated 25 times, and the evolution of the loss function was recorded.
Each training run was performed for 30{,}000 epochs, and the loss value was recorded every 100 epochs. The resulting loss curves are shown in Figure~\ref{fig:loss_curve}. In the figure, thin and light lines represent the loss curves for individual runs, while thick and dark lines represent the mean loss curves over the 25 trials.
As shown in the figure, the second stage with transfer learning exhibits faster reduction in the loss value during the early and middle phases of training (up to approximately 15{,}000 epochs), reaching the convergence region (loss between $10^{-1}$ and $10^{-2}$) significantly earlier than the first stage. Moreover, the initial loss value at epoch 0 is also substantially lower in the second stage, indicating the advantage of using pre-trained weights.
On the other hand, in situations where higher accuracy is required (e.g., loss below $10^{-3}$), the benefit of transfer learning becomes less evident in the later phase of training, and its impact on computational efficiency may diminish.
Therefore, this approach is particularly effective when the required precision is moderate, enabling efficient training without sacrificing performance within practical accuracy constraints.

\section{Conclusions}\label{sec3}
In this study, the compressible Neural Particle Method (cNPM) is proposed for simulating compressible viscous fluid flows.
The method incorporates the Tait equation of state to account for density variations in weakly compressible regimes. 
To validate its performance, cNPM is applied to a dam-break scenario, and the results are quantitatively compared with those obtained using Smoothed Particle Hydrodynamics (SPH).
The cNPM demonstrates enhanced numerical stability with respect to irregular particle distributions and yields pressure time-series that closely match those of SPH. 
Furthermore, the introduction of transfer learning reduces computational time, making the method more efficient for practical applications. 
In future research, we will investigate the range of compressibility levels that cNPM can handle. 
Specifically, we will focus on problem settings where the effects of compressibility are more pronounced, such as increasing water depth and using fluids with higher compressibility than water, within a model employing the Tait equation. 
Moreover, to extend the range of states that can be simulated, we will incorporate an energy equation into the governing equations and adopt the Tillotson equation of state, which can represent a wider range of materials, pressures, and temperatures.

\section{Appendix : Viscous Term Representation in the Update Lagrangian Formulation}

The viscous term of the Navier--Stokes equations is expressed using the update Lagrangian formulation. For the introduction of update lagrange formulation to NPM, please refer to \cite{Wessels2020}. The $k$-stage approximation of the Runge--Kutta method is given by
\begin{equation}
  \begin{aligned}
  \bm{x}^j & =\bm{x}_n+\Delta t \sum_{i=1}^k a^{j i} \bm{v}^i, \\
  \bm{x}_{n+1} & =\bm{x}_n+\Delta t \sum_{j=1}^k b^j \bm{v}^j.
  \end{aligned}
\end{equation}

We introduce the incremental deformation gradient tensor, which expresses the partial derivative of the velocity with respect to the stage position as
\begin{equation}
  \begin{aligned}
  \Delta \dot{\bm{F}}^i &= \frac{\partial \bm{v}^i}{\partial \bm{x}_n}, &
  \Delta \bm{F}^i &= \frac{\partial \bm{x}^i}{\partial \bm{x}_n} = \bm{1} + \Delta t \sum_{i=1}^k a^{j i} \frac{\partial \bm{v}^i}{\partial \bm{x}_n} = \bm{1} + \Delta t \sum_{i=1}^k a^{j i} \Delta \dot{\bm{F}}^i, \\
  \end{aligned}
\end{equation}

Next, the Laplacian of the velocity output from the neural network, which corresponds to the viscous term, is computed. By definition, the Laplacian of the velocity is given as
\begin{equation}
\begin{aligned}
\nabla \cdot \nabla \bm{v}^i 
&= \mathrm{tr}\left(\frac{\partial}{\partial \bm{x}^i} \left(\frac{\partial \bm{v}^i}{\partial \bm{x}^i}\right)\right) \\
&= \mathrm{tr}\left(\frac{\partial}{\partial \bm{x}_n} \left(\frac{\partial \bm{v}^i}{\partial \bm{x}^i}\right)\frac{\partial \bm{x}_n}{\partial \bm{x}^i}\right) \\
&= \mathrm{tr}\left(\frac{\partial}{\partial \bm{x}_n} \left(\frac{\partial \bm{v}^i}{\partial \bm{x}_n} \frac{\partial \bm{x}_n}{\partial \bm{x}^i}\right)\Delta (\bm{F}^i)^{-1}\right) \\
&= \mathrm{tr}\left(\frac{\partial}{\partial \bm{x}_n} \left(\Delta \dot{\bm{F}^i} \Delta (\bm{F}^i)^{-1}\right)\Delta (\bm{F}^i)^{-1}\right).
\end{aligned}
\end{equation}

In general, the second-order derivative of a vector is expressed using the Hessian matrix as
\begin{equation}
  \begin{aligned}
  \nabla \cdot \nabla \bm{v}
  &= \left(\mathrm{tr} \nabla^2 v_x, \mathrm{tr} \nabla^2 v_y\right) \\
  &= \left(\mathrm{tr} \left(\frac{\partial}{\partial \bm{x}}
  \left(\frac{\partial v_x}{\partial x}, \frac{\partial v_x}{\partial y}\right)\right), 
    \mathrm{tr} \left(\frac{\partial}{\partial \bm{x}}
  \left(\frac{\partial v_y}{\partial x}, \frac{\partial v_y}{\partial y}\right)\right)\right)\\
  &= \left(
    \frac{\partial^2 v_x}{\partial x^2} + \frac{\partial^2 v_x}{\partial y^2}, \frac{\partial^2 v_y}{\partial x^2} + \frac{\partial^2 v_y}{\partial y^2}
  \right)
  \end{aligned}
\end{equation}

By defining $\bm{l} = \Delta \dot{\bm{F}} \Delta \bm{F}^{-1}$, the above expression can be rewritten as
\begin{equation}
  \begin{aligned}
  \nabla \cdot \nabla\bm{v}^i &= \mathrm{tr} \left(\frac{\partial}{\partial \bm{x}_n}
  \left(\begin{array}{ll}
    l_{11}& l_{12}\\
    l_{21}& l_{22}
    \end{array}\right)
    \Delta (\bm{F}^i)^{-1}\right)\\
  &= \left(
    \frac{\partial l_{11}}{\partial x_n} \Delta F^{-1}_{11} + \frac{\partial l_{11}}{\partial y_n} \Delta F^{-1}_{21} + \frac{\partial l_{12}}{\partial x_n} \Delta F^{-1}_{12} + \frac{\partial l_{12}}{\partial y_n} \Delta F^{-1}_{22},
    \right. \\
    &\quad\quad
    \left.
    \frac{\partial l_{21}}{\partial x_n} \Delta F^{-1}_{11} + \frac{\partial l_{21}}{\partial y_n} \Delta F^{-1}_{21} + \frac{\partial l_{22}}{\partial x_n} \Delta F^{-1}_{12} + \frac{\partial l_{22}}{\partial y_n} \Delta F^{-1}_{22}
    \right).
  \end{aligned}
\end{equation}

Since $\bm{l}$ is computed from the output of the neural network, the Laplacian of the velocity can be easily computed by applying automatic differentiation with respect to the neural network inputs $x_n, y_n$.

\section*{Data availability}
The source code for the numerical examples is available for download from \url{https://doi.org/10.5281/zenodo.16310224}.

\section*{Acknowledgments}
The numerical implementation used in this study was developed based on the code publicly available on GitLab (\url{https://gitlab.com/henningwessels/npm}) by the authors of the Wessels paper~\cite{Wessels2020}. Their open-source contribution was invaluable for enabling the computations presented in this work and provided crucial guidance for the implementation.  
We also express our sincere gratitude for their kind responses to technical inquiries, which significantly enhanced our understanding. This work was supported by the Heiwa Nakajima Foundation.

\section*{Declaration of generative AI and AI-assisted technologies in the writing process}
During the preparation of this work, the author Masato Shibukawa used ChatGPT in order to improve the language, grammar, and readability of the manuscript. After using this tool, the author reviewed and edited the content as needed and takes full responsibility for the content of the publication.




\bibliography{main.bib} 

\begin{thebibliography}{10}

\bibitem{Lind2020}
S.~J. Lind, B.~D. Rogers, and P.~K. Stansby.
\newblock Review of smoothed particle hydrodynamics: towards converged lagrangian flow modelling.
\newblock {\em Proc. R. Soc. A}, 476:20190801, 2020.

\bibitem{Shadloo2016}
M.~S. Shadloo, G.~Oger, and D.~Le Touz{\'e}.
\newblock Smoothed particle hydrodynamics method for fluid flows, towards industrial applications: Motivations, current state, and challenges.
\newblock {\em Computers and Fluids}, 136:11--34, 2016.

\bibitem{Ye2019}
T.~Ye, D.~Y. Pan, C.~Huang, and M.~B. Liu.
\newblock Smoothed particle hydrodynamics (sph) for complex fluid flows: Recent developments in methodology and applications.
\newblock {\em Phys. Fluids}, 31, 2019.

\bibitem{lind2012}
S.~J. Lind, R.~Xu, P.~K. Stansby, and B.~D. Rogers.
\newblock Incompressible smoothed particle hydrodynamics for free-surface flows: A generalised diffusion-based algorithm for stability and validations for impulsive flows and propagating waves.
\newblock {\em J. Comput. Phys.}, 231:1499--1523, 2012.

\bibitem{xu2009}
R.~Xu, P.~Stansby, and D.~Laurence.
\newblock Accuracy and stability in incompressible sph (isph).
\newblock {\em J. Comput. Phys.}, 228(18):6703--6720, 2009.

\bibitem{sun2012}
P.~N. Sun, A.~Colagrossi, S.~Marrone, M.~Antuono, and A.-M. Zhang.
\newblock A consistent approach to particle shifting in the $\delta$-plus-sph model.
\newblock {\em Computer Methods Appl. Mech. Eng.}, 348:912--934, 2019.

\bibitem{vacondio2013}
R.~Vacondio, B.~Rogers, P.~K. Stansby, P.~Mignosa, and J.~Feldman.
\newblock Variable resolution for sph: A dynamic particle coalescing and splitting scheme.
\newblock {\em Computer Methods Appl. Mech. Eng.}, 256:132--148, 2013.

\bibitem{molteni2009}
D.~Molteni and A.~Colagrossi.
\newblock A simple procedure to improve the pressure evaluation in hydrodynamic context using the sph.
\newblock {\em Comput. Phys. Commun.}, 180:861--872, 2009.

\bibitem{colagrossi2003}
A.~Colagrossi and M.~Landrini.
\newblock Numerical simulation of interfacial flows by smoothed particle hydrodynamics.
\newblock {\em J. Comput. Phys.}, 191:448--475, 2003.

\bibitem{marrone2011}
S.~Marrone, M.~Antuono, A.~Colagrossi, G.~Colicchio, D.~Le Touz{\'e}, and G.~Graziani.
\newblock $\delta$-sph model for simulating violent impact flows.
\newblock {\em Computer Methods Appl. Mech. Eng.}, 200:1526--1542, 2011.

\bibitem{hopkins2015}
P.~F. Hopkins.
\newblock A new class of accurate, mesh-free hydrodynamic simulation methods.
\newblock {\em Mon. Not. R. Astron. Soc.}, 450(1):53--110, 2015.

\bibitem{Kutz2017}
J.~N. Kutz.
\newblock Deep learning in fluid dynamics.
\newblock {\em J. Fluid Mech.}, 814:1--4, 2017.

\bibitem{Kochkov2021}
D.~Kochkov, J.~A. Smith, A.~Alieva, Q.~Wang, M.~P. Brenner, and S.~Hoyer.
\newblock Machine learning--accelerated computational fluid dynamics.
\newblock {\em Proc. Natl. Acad. Sci. USA}, 118:e2101784118, 2021.

\bibitem{Raissi2019}
M.~Raissi, P.~Perdikaris, and G.~E. Karniadakis.
\newblock Physics-informed neural networks: A deep learning framework for solving forward and inverse problems involving nonlinear partial differential equations.
\newblock {\em J. Comput. Phys.}, 378:686--707, 2019.

\bibitem{Cai2022}
S.~Cai, Z.~Mao, Z.~Wang, M.~Yin, and G.~E. Karniadakis.
\newblock Physics-informed neural networks (pinns) for fluid mechanics: A review.
\newblock {\em Acta Mechanica Sinica}, 37:1--12, 2022.

\bibitem{Mao2020}
Z.~Mao, A.~D. Jagtap, and G.~E. Karniadakis.
\newblock Physics-informed neural networks for high-speed flows.
\newblock {\em Comput. Methods Appl. Mech. Engrg.}, 360, 2020.

\bibitem{Ko2025}
Seungchan Ko and Sanghyeon Park.
\newblock Vs-pinn: A fast and efficient training of physics-informed neural networks using variable-scaling methods for solving pdes with stiff behavior.
\newblock {\em Journal of Computational Physics}, 529:113860, 2025.

\bibitem{Yadav2024}
Vikas Yadav, Mario Casel, and Abdulla Ghani.
\newblock Rf-pinns: Reactive flow physics-informed neural networks for field reconstruction of laminar and turbulent flames using sparse data.
\newblock {\em Journal of Computational Physics}, 524:113698, 2025.

\bibitem{Jin2021}
X.~Jin, S.~Cai, H.~Li, and G.~E. Karniadakis.
\newblock Nsfnets (navier--stokes flow nets): Physics-informed neural networks for the incompressible navier--stokes equations.
\newblock {\em J. Comput. Phys.}, 426, 2021.

\bibitem{Raissi2020}
M.~Raissi, A.~Yazdani, and G.~E. Karniadakis.
\newblock Hidden fluid mechanics: Learning velocity and pressure fields from flow visualizations.
\newblock {\em Science}, 367:1026--1030, 2020.

\bibitem{Haghighat2021}
E.~Haghighat, M.~Raissi, A.~Moure, H.~Gomez, and R.~Juanes.
\newblock A physics-informed deep learning framework for inversion and surrogate modeling in solid mechanics.
\newblock {\em Comput. Methods Appl. Mech. Engrg.}, 379, 2021.

\bibitem{Wessels2020}
H.~Wessels, C.~Wei{\ss}enfels, and P.~Wriggers.
\newblock The neural particle method---an updated lagrangian physics informed neural network for computational fluid dynamics.
\newblock {\em Comput. Methods Appl. Mech. Engrg.}, 368, 2020.

\bibitem{Bai2022}
J.~Bai, Y.~Zhou, Y.~Ma, H.~Jeong, H.~Zhan, C.~Rathnayaka, E.~Sauret, and Y.~Gu.
\newblock A general neural particle method for hydrodynamics modeling.
\newblock {\em Comput. Methods Appl. Mech. Engrg.}, 393:114740, 2022.

\bibitem{Becker2007}
M.~Becker and M.~Teschner.
\newblock Weakly compressible sph for free surface flows.
\newblock In {\em Proc. ACM SIGGRAPH/Eurographics Symposium on Computer Animation (SCA '07)}, pages 209--217, 2007.

\bibitem{Monaghan1994}
J.~Monaghan.
\newblock Simulating free surface flows with sph.
\newblock {\em J. Comput. Phys.}, 110(2):399--406, 1994.

\bibitem{Schmidhuber2015}
J.~Schmidhuber.
\newblock Deep learning in neural networks: An overview.
\newblock {\em Neural Networks}, 61:85--117, 2015.

\bibitem{Baydin2018}
A.~G. Baydin, B.~A. Pearlmutter, A.~A. Radul, and J.~M. Siskind.
\newblock Automatic differentiation in machine learning: a survey.
\newblock {\em Journal of Machine Learning Research}, 18, 2018.

\bibitem{Iserles2009}
A.~Iserles.
\newblock {\em A First Course in the Numerical Analysis of Differential Equations}.
\newblock Cambridge University Press, 2009.

\bibitem{Zhuang2021}
F.~Zhuang et~al.
\newblock A comprehensive survey on transfer learning.
\newblock {\em Proc. IEEE}, 109(1):43--76, 2021.

\bibitem{Misra2019}
D.~Misra.
\newblock Mish: A self regularized non-monotonic neural activation function.
\newblock \url{https://arxiv.org/abs/1908.08681}, 2019.
\newblock arXiv preprint arXiv:1908.08681.

\end{thebibliography}
\bibliographystyle{unsrt} 
\end{document}